\documentclass[12pt]{article}
\begin{document}
\begin{center}
{\bf{ f(R) Gravity with  k-essence scaling relation and Cosmic acceleration
}}

\bigskip

{\bf{
Debashis Gangopadhyay$^\ast$ and Somnath Mukherjee$^\dagger$

\medskip
{$\ast$Department of Physics, Ramakrishna Mission Vivekananda University, Belur Matth, Howrah-711202, India.\\
debashis@rkmvu.ac.in\\

$^\dagger$Dharsa Mihirlal Khan Institution[H.S],
            P.O:-New G.I.P Colony,  Dist:-Howrah-711112, India.\\
sompresi@gmail.com}}}
\end{center}
 
\begin{center}
Abstract  
\end{center} 
A modified gravity theory with $f(R)=R^2$ coupled to a dark energy lagrangian  $L=-V(\phi)F(X)$ , $X=\nabla_{\mu}\phi\nabla^{\mu}\phi$, gives  
plausible cosmological scenarios when the modified Friedman equations are solved  subject to the scaling relation $X (\frac{dF}{dX})^{2}=Ca(t)^{-6}$. This relation is already known to be valid, for constant potential $V(\phi)$, when $L$ is coupled to Einstein gravity. $\phi$ is the k-essence scalar field and $a(t)$ is the scale factor. The various scenarios are: 
(1) Radiation dominated Ricci flat universe with deceleration parameter $Q=1$. The solution for $\phi$ is an inflaton field  for small times. (2) $Q$ is always negative and we have accelerated expansion of the universe right from the beginning of time and $\phi$ is an  inflaton for small times. (3)The deceleration parameter $Q= -5$, i.e. we have an accelerated expansion of the universe. $\phi$ is an inflaton for small times.(4)A generalisation to $f(R)= R^n$ shows that whenever $n > 1.780$ or $n < - 0.280$ , $Q$ will be negative and we will have accelerated expansion of the universe. At small times $\phi$ is again an inflaton.

The results remind us of other physical phenomena where existence 
of scaling relations signal some sort of universality for theories with different 
microscopic lagrangians. Here this is seen in the case of Einstein gravity and modified gravity theories

\section{Introduction}

Very often the existence of similar scaling relations 
in apparently different  physical systems signify the presence of 
some sort of universality . A physical system 
is usually typified by some lagrangian from which an action is constructed.
Therefore, different actions describe different theories. However, if the  
same scaling relation is valid in theories with different actions, then it is 
possible that these theories may describe physical scenarios which have 
certain aspects in common. The motive of this paper is to show that in cosmology 
such a situation can occur  in the context of dark  energy where the dark energy 
is realised  by $k-$ essence scalar fields $\phi$. 

A $k-$essence theory coupled with a non-canonical lagrangian coupled to Einstein gravity is known to satisfy a scaling 
relation. Consider the usual Einstein-Hilbert action $S_{\rm EH}= \frac{1}{2\kappa}\int d^4 x \sqrt{-g}\,R$ in presence of  $k-$essence fields $\phi$ with a non-canonical lagrangian 
$L=-V(\phi)F(X)$, $X=\nabla_{\mu}\phi\nabla^{\mu}\phi$. 
$\kappa\equiv 8\pi G$, $G$ is the gravitational constant, 
$g$ is the determinant of the metric and we work in units where $c=\hbar=1$. Relevant 
literature on dark energy and $k-$essence can be found in 
\cite{R1,R2,R3,R4,R5,R6,R7,R8,R9,R10,R11,R12,R13,R14}.

For constant potential $V(\phi)$, it follows from the 
resulting Friedman equations that $X (\frac{dF}{dX})^{2}=Ca(t)^{-6}$  \cite{R15,R16,R17,R18}. In this work we investigate what happens if this same 
scaling relation is required to be valid when the same $k-$essence lagrangian is coupled to modified gravity or $f(R)$ gravity theories where $f$ is a general function of the Ricci scalar $R$ \cite{R19,R20,R21,R22,R23,R24,R25,R26,R27,R28,R29,R30,R31,R32}.  

There are various ways of obtaining such  theories. The modified Friedman equations 
for such $f(R)$ gravity theories were first obtained in \cite{R19}.
We shall follow the notations of \cite{R23},\cite{R29}.
In one approach, $f(R)$ theories of gravity are obtained by generalising 
the Einstein-Hilbert action into 
\begin{equation}
S_{\rm f} =\frac{1}{2\kappa} \int d^4 x \sqrt{-g} \, f(R)
\end{equation}
where $f(R)= R$ gives the usual Einstein gravity. 

We work in a Freidman-Lemaitre-Robertson-Walker metric with curvature 
constant $k=0$ : 
\begin{equation}
ds^{2}=c^{2}dt^{2}-a^{2}(t)[dr^{2} 
+r^{2}(d\theta^{2}+sin^{2}\theta d\phi^{2})]
\end{equation}
Adding a matter term 
$S_{\rm M}$, the total action   
\begin{equation} 
S=\frac{1}{2\kappa}\int d^4 x \sqrt{-g} \, f(R) 
+S_{\rm M}(g_{\mu\nu},\phi)
\end{equation}
where $\phi$ are generic matter fields. We shall consider a single 
scalar field only. The field equations are \cite{R23},\cite{R29}:
\begin{equation}
 f'(R)R_{\mu\nu}-\frac{1}{2}f(R)g_{\mu\nu}- 
\left[\nabla_\mu\nabla_\nu -g_{\mu\nu} \nabla^\beta\nabla_\beta\right] f'(R)\nonumber\\
 = \kappa \,T_{\mu\nu} 
\end{equation}
with  
\begin{equation}
T_{\mu\nu}=\frac{-2}{\sqrt{-g}}\, \frac{\delta
S_M }{\delta g^{\mu\nu} }
\end{equation}
A prime denotes differentiation with respect to the argument. 
$\nabla_\mu$ is the covariant derivative associated with the 
Levi-Civita connection of the metric. The $00-$ component of the field  
equations (4) gives the modified Friedmann equations:
\begin{equation}
 ({\frac{\dot {a}}{a}})^{2}- {\frac{1}{3f'(R)}}\biggl({\frac{1}{2}}[f(R)-Rf'(R)] 
-3({\frac{\dot {a}}{a}})\dot R f''(R)\biggr)\nonumber\\
 = \frac{1}{3} \kappa\rho 
\end{equation}
while the $ii$ components give 
\begin{equation}
 2(\frac{\ddot a}{a})+ (\frac{\dot a}{a})^{2}
+\frac{1}{f'(R)}\biggl(2(\frac{\dot a}{a})\dot {R}f''(R)+\ddot{R}f''(R)\nonumber\\
+(\dot R)^{2}f'''(R) -\frac{1}{2}[f(R)-Rf'(R)]\biggr)=-\kappa p
\end{equation}
We first recall the relevant equations for the usual case $f(R)=R$ \cite{R15,R16,R17,R18}.
The energy density obtained from (6) is 
\begin{equation}
\rho= \frac {3}{\kappa}  H^{2} 
\end{equation}
while (7) gives the pressure as 
\begin{equation}
p= -\frac{2\ddot a}{\kappa a} - \frac{H^{2}}{\kappa}
\end{equation}
Differentiating (8) with respect to time, and using (9)
gives 
\begin{equation}
\dot\rho + 3H(\rho + p) = 0
\end{equation}
Now, as already stated before, the lagrangian of $k-$essence scalar fields $\phi$ is non-canonical and is of the form  
${\mathcal L}=p= -V(\phi) F(X)$
where $V(\phi) $ is the potential.
The energy density is $\rho = V(\phi)[ F(X) -2 X F_{X}]$
with $F_{X}\equiv\frac {dF}{dX}$. Substituting these values of $\rho$ and 
$p$  in (10) and taking $V(\phi)=constant$ one gets 
\begin{equation}
(\frac {dF}{dX} + 2 X \frac{d^{2}F}{dX^{2}}) a \frac {dX}{da}
+6 \frac{dF}{dX} X = 0 
\end{equation}
This equation can be integrated to give the  scaling relation \cite{R15,R16,R17,R18}  
\begin{equation}
X (\frac{dF}{dX})^{2}=Ca(t)^{-6} 
\end{equation}
$C$ is a constant. 

Using (12) and 
the zero-zero component of Einstein's field equations an expression for the lagrangian 
for the $k-$essence field can be obtained for a FLRW metric. This has been elaborately described in \cite{R17} where the scaling relation (12) has been used to eliminate $F_{X}$. For a homogeneous scalar field $\phi(t)$ the lagrangian is \cite{R17}
\begin{equation}
L=-c_{1}\dot q^{2} - c_{2} V(\phi)\dot \phi e^{-3q}
\end{equation}
where $q(t)=ln\enskip a(t)$, $c_{1}= 3(8\pi G)^{-1}$, $c_{2}=2 \sqrt C$.
The new lagrangian has two generalised coordinates $q(t)$ and $\phi(t)$.
$q$ has a standard kinetic term while $\phi$ does not have a kinetic part.
There is a complicated polynomial interaction between $q$ and $\phi$ and $\phi$ 
occurs purely through this interaction term.

We now consider the case for $f(R)=R^{2}$. Then  (6) gives for the energy density 
\begin{equation}
\rho = \frac {3}{\kappa}  H^{2} + \frac {R}{4\kappa} + \frac {3H}{\kappa}\frac {\dot R}{R}
\end{equation}
while (7) gives the pressure ( or lagrangian)  
\begin{equation}
p= -\frac {2\ddot {a}}{\kappa a} - \frac {H^{2}}{\kappa}-\frac {2H\dot R}{\kappa R}-\frac {\ddot {R}}{\kappa R} - 
\frac {R}{4\kappa} 
\end{equation}
Using $\frac {\ddot a}{a}=\dot H + H^{2}$ we have 
\begin{equation}
\dot\rho + 3H(\rho + p) = \frac {\dot R}{4\kappa}+ 3\dot H\frac {\dot R}{\kappa R}-3H\frac {\dot R ^{2}}{\kappa R^{2}}
+3H^{2}\frac {\dot R}{\kappa R}
\end{equation}
It is readily seen that (16) reduces to (10) 
if the right hand side vanishes i.e.
\begin{equation}
\dot R [R^{2} + 12 \dot H R -12 H \dot R +12 H^{2} R] = 0
\end{equation}
i.e. if either $\dot {R} =0$ or $R^{2} + 12 \dot {H} R -12 H \dot {R} +12 H^{2} R = 0$. 
Now $R=6[\dot {H} + 2 H^{2}]$ so that $\dot R = 6[\ddot {H} + 4 H \dot {H}]$.
Using  this in (17) we have two scenarios for which (16) reduces to (10), {\it viz.}: 
\begin{equation}
\ddot{ H} + 4 H \dot{ H}=0
\end{equation}
and
\begin{equation}
3{\dot{ H}}^{2} - 2 H \ddot{ H} + 2 H^{2} \dot{ H} + 8 H^{4}=0
\end{equation}

\section{Solutions for Hubble parameter}

Now we know the conditions (i.e. equations (18) or (19))
that the Hubble parameter has to satisfy so that 
the scaling relation (12) is valid in a modified gravity theory with 
$f(R)=R^{2}$. 

\subsection{Solutions for $\dot R=0$}

A first integral of equation (18) gives $R=6(\dot H + 2H^2) = 6A$ where $A$ is a constant.

{\it Case 1: $A=0$ , i.e.$R=0$ (Ricci flat):}

If the constant $A$ is chosen to be zero then we have $R=6[\dot {H} + 2 H^{2}]=0$
so that the emergent spacetime is Ricci flat. A simple integration gives $H=\frac{1}{2t}$.
This implies that the scale factor is $a(t)\sim t^{\frac{1}{2}}$ and the cosmology is that of  
the well known  radiation dominated universe.
From (14), the energy density is 
\begin{equation}
\rho= \frac {3}{4\kappa t^{2}}
\end{equation}
because in this case $R= 6 [\dot {H} + 2H^{2}]=0$. The lagrangian or pressure 
can be obtained from (15) as 
\begin{equation}
p= \frac {1}{4\kappa t^{2}}
\end{equation}
To determine the $k-$essence scalar field $\phi$ recall \cite{R17}. 
Take $V(\phi)=V= constant$ then one has  
\begin{equation}
 L=p=-V(\phi)F(X)=-c_{1}(\frac{\dot a}{a})^{2}-c_{2}V\dot\phi a^{-3}\nonumber\\
 =c_{1}(\frac{1}{4t^{2}})-c_{2}V\dot\phi t^{-3/2}
\end{equation}
where $c_{1}= \frac{3}{\kappa}$ and $c_{2}=2{\sqrt C}$. It is to be noted 
that the equation (22) incorporates the scaling relation.  
Equations (21) and (22) then give the solution for the $k-$essence scalar field as 
\begin{equation}
\phi(t)=\frac {1}{32\pi G{\sqrt C}V} t^{\frac {1}{2}}
\end{equation}
where we have put all integration constants to zero. Now $t\equiv\frac{t}{t_{0}}$ 
where $t_{0}$ is the present epoch and so $t$ is always less than unity. 
Now for $0 < t < 2$, $ln~t\sim (t-1)$. 
So $t^{1/2} = e^{\frac{1}{2}~ln~t}\sim e^{\frac{1}{2}(t-1)}\sim 1 + t$. 
Hence for small times $\phi(t)\sim \frac {1}{64\pi G{\sqrt C}V} + \frac {1}{64\pi G{\sqrt C}V} t$ and 
this is like the scalar field in "chaotic inflation" as in \cite{R33}. 
  
The deceleration parameter $Q=-\frac{a\ddot a}{{\dot{a}}^{2}}=1$ while the equation of state 
parameter $w=\frac {p}{\rho}=\frac {1}{3}$. So we have a  Ricci flat ($R=0$)
decelerating universe with radiation domination.

The interesting aspect is that this has been obtained from a modified gravity 
theory having a {\it dark energy constituent satisfying a scaling relation that is 
also satisfied in Einstein  gravity.}

{\it Case 2: $A\neq 0$ , i.e. $R\neq 0$ :}

If $A\neq 0$, then a solution for the Hubble parameter is (choosing an integration constant to be zero) 
\begin{equation}
H(t)=\sqrt{\frac{A}{2}}\tanh{[\sqrt{2A} t]}
\end{equation}
The scale factor is then
\begin{equation}
a(t)=\cosh^{\frac{1}{2}}{[\sqrt{2A} t]}
\end{equation}
and the deceleration parameter is  
\begin{equation}
Q = 1 - 2\coth^{2}{ [\sqrt {2A} t]}
\end{equation}
Hence the deceleration parameter is always negative as $1\leq \coth t\leq\infty$ for $t\geq 0$.
So we have accelerated expansion of the universe right from the beginning of time.

Pressure $p$ now is 
\begin{equation}
p=-\frac{1}{\kappa}[2\dot{H}+3H^{2}+\frac{R}{4}]
\end{equation}
Putting in $R=6A$ and $2\dot{H}=2A-4H^{2}$ gives 
\begin{equation}
p=\frac{H^{2}}{\kappa}-\frac{7A}{2\kappa}
=\frac{A}{2\kappa}\tanh^{2}[\sqrt{2A}t]-\frac{7A}{2\kappa}
\end{equation}
where we have put in $H$ from (24).

Alternatively, using (22) we have  
\begin{equation}
p=-c_{1}\frac{A}{2}\tanh^{2}[\sqrt{2A}t]-c_{2}V\dot{\phi}\cosh^{-\frac{3}{2}}[\sqrt{2A}t]
\end{equation}
Equating (28) and (29) we get 
$\frac{d\phi}{dt}=\frac{7A}{2\kappa c_{2}V}\cosh^{\frac{3}{2}}[\sqrt{2A}t]
-\frac{A}{2c_{2}V}(\frac{1}{\kappa}+c_{1})\tanh^{2}[\sqrt{2A}t]\cosh^{\frac{3}{2}}[\sqrt{2A}t]$
which upon integration gives 
\begin{equation}
\phi(t)= \bigg(\frac{\sqrt{A}}{\sqrt{2}\kappa c_{2}V}\bigg)\sinh[\sqrt{2A}t]\cosh^{\frac{1}{2}}[\sqrt{2A}t]\nonumber\\
 -\bigg(\frac{15\sqrt{A}}{\sqrt{2}\kappa c_{2}V}\bigg)i\frac{F(\frac{i\sqrt{2A}t}{2}|2)}{3} + constant
\end{equation}
where the elliptic function $F$ is defined as $F(\alpha | m)=\int_{0}^{\alpha} (1-m sin^{2}\theta)^{-\frac{1}{2}}d\theta$.  Note  that  for $\alpha\rightarrow 0$, $F\rightarrow 0$. Therefore, in the early 
universe where $t\equiv \frac{t}{t_{0}} << 1$, $t_{0}$ being the present epoch, the $F$ term in (30) can be safely ignored. Then for early times $\phi(t)$ again qualifies for an inflationary field \cite{R33} since 
the equation (30) gives $\phi(t)\sim const. + \bigg(\frac{A}{\kappa c_{2}V}\bigg)~t$.

\subsection{Solutions for $\dot R\neq 0$}

If $\dot R\neq 0$, then (19) should hold.    
The equation (19) is readily solved with the ansatz $H=\frac{\alpha}{t}$ with 
$\alpha$ a constant. Using this ansatz ,(19) gives for $\alpha\neq 0$ 
\begin{equation}
8\alpha ^{2} - 2 \alpha - 1 = 0
\end{equation}
This is a quadratic in $\alpha$ and for real $\alpha$ the  solutions are 
$\alpha=\frac {1}{2},   -\frac {1}{4}$ so that the Hubble parameter 
solutions corresponding to the two values of $\alpha$ are 
\begin{equation}
H =\frac {1}{2t}~~;~~ -\frac {1}{4t}
\end{equation}
Of these two solutions , we have already encountered the first one.
So we consider the other solution only, i.e. $H= - \frac{1}{4t}$. This 
gives the solution for the scale parameter as 
\begin{equation}
a(t)= const. t^{-\frac {1}{4}}
\end{equation}
The deceleration parameter is now $Q= -5$. So again we have an 
accelerated expansion owing its origin to dark energy.  

The Ricci scalar is 
$R=6[\dot{ H} + 2H^{2}]= \frac{9}{4t^{2}}$.
The energy density using $(14)$  now is  
\begin{equation}
\rho = \frac {9}{4\kappa t^{2}}
\end{equation}
while  from  (15) the pressure is obtained as 
\begin{equation}
p = -\frac{33}{4\kappa t^{2}}
\end{equation}
As before, equating (35) to the lagrangian gives
\begin{equation}
-\frac{33}{4\kappa t^{2}}=-c_{1}(\frac {\dot a}{a})^{2}-c_{2}V\dot\phi a^{-3}\nonumber\\
= - c_{1}(\frac {1}{16t^{2}})-c_{2}V\dot\phi (const.) t^{-3/4}
\end{equation}
Proceeding as before, a solution for the dark energy scalar field is obtained as 
\begin{equation}
\phi(t)= (const.) t^{-\frac{1}{4}}
\end{equation}
Note that we can write 
$\phi(t)\equiv (const.) e^{-\frac{1}{4}ln t}\sim e^{-\frac{1}{4}(t-1)}$ for $0 < t < 2$. 
In our case this is always true as $t\equiv\frac{t}{t_{0}}$ always lies between $0$ and $1$.
Therefore
$\phi(t)\sim e^{-\frac{1}{4}(t-1)}= 1 -\frac{1}{4}t$ again qualifies for an inflationary field \cite{R33}.

\section{Generalisation to $f(R)= R^{ n}$}

We shall now consider  the general case for $f(R)=R^{n}$ and for convenience 
take $\kappa=1$.  
Then $f'(R)=nR^{(n-1)}$ ; $f''(R)=n(n-1)R^{(n-2)}$ ; $f'''(R)=n(n-1)(n-2)R^{(n-3)}$ 
and equation (4)  becomes  
\begin{equation}
{\rho={3H^2}+{(n-1)\over{2n}}R+{3(n-1)\dot{R}\over{R}}H}
\end{equation}
so that 
\begin{equation}
\dot\rho=6H\dot{H}+ \frac{n-1}{2n}\dot{R} + \frac{3(n-1)}{R}\dot R\dot{H}\nonumber\\
+ \frac{3(n-1)\ddot R}{R} H -\frac{3(n-1)\dot{R}^2}{R^2}H
\end{equation}
The pressure density is given by 
\begin{equation}
p =-2\dot H - 3H^2 - \frac{(n-1)(2H\dot R + \ddot R)}{R}\nonumber\\
- \frac{(n-1)(n-2)\dot R^2}{R^2} - \frac{(n-1)}{2n} R
\end{equation}
From (38) and (40) we have 
\begin{equation}
3H(\rho + p)= -6H\dot H +3(n-1)\frac{\dot R}{R}H^2\nonumber\\
- 3(n-1)\frac{\ddot R}{R}H 
- 3(n-1)(n-2)\frac{\dot R ^2}{R^2} H
\end{equation}
Using (39) and (41) gives the equation of continuity 
\begin{equation}
\dot{\rho}+3H(\rho+P)={{(n-1)\over{2n}}\dot{R}}\nonumber\\
+{3(n-1)}{\dot{R}\over{R}}[{\dot{H}}+{H^2}-{(n-1){\dot{R}\over{R}}H}]
\end{equation}
For the scaling relation (12)  to be valid, the right hand side of (42) must vanish. This gives
(note that $n\neq 1$)
\begin{equation}
\dot R [\frac{1}{2n}+ \frac{3}{R}\big(\dot H + H^2- \frac{(n-1)\dot R H}{R}\big)]=0
\end{equation}
There are two possibilities: 

First $\dot{R}=6[\ddot{H}+4H\dot{H}]=0$ and $[\frac{1}{2n}+ \frac{3}{R}\big(\dot H + H^2- \frac{(n-1)\dot R H}{R}\big)]\neq0$. This gives a solution 
$H(t) = \frac{1}{2t}$. This we have already encountered and we have a radiation dominated universe.

The other solution is obtained when 
$\dot{R}\neq 0$. Then 
\begin{equation}
[\frac{1}{2n}+ \frac{3}{R}\big(\dot H + H^2- \frac{(n-1)\dot R H}{R}\big)]=0
\end{equation}
Substituting $R=6[\dot{H}+2H^2]$ and $\dot{R}=6[\ddot{H}+4H\dot{H}]$
in (44) and solving for $H$ we get 
\begin{equation}
H= \frac{(-2n^{2}+3n+1)}{(2+n)t} = \frac{u}{t}
\end{equation}
where $u=\frac{(-2n^{2}+3n+1)}{(2+n)}$.

So the scale factor is obtained as 
\begin{equation}
a(t)= t^{u}
\end{equation}

The deceleration parameter $Q=-[1+\frac{\dot H}{H^2}]$ now is  
\begin{equation}
Q=-[1-\frac{(2+n)}{(-2n^{2}+3n+1)}]
\end{equation}
For late time acceleration of the universe $Q$ should always be negative i.e.
the term within third brackets in (47) must be positive i.e.  
$1-\frac{(2+n)}{(-2n^{2}+3n+1)} > 0$ or $\frac{(2+n)}{(-2n^{2}+3n+1)}<1$. 
This means that whenever 
$n < \frac{1}{4}(3-\sqrt{17})= - 0.280$ or $n > \frac{1}{4}(3+\sqrt{17}) = 1.780$ there will be accelerated expansion  of the universe. 

We now determine the $k-$essence scalar field $\phi$.
The lagrangian or pressure is 
\begin{equation}
p=-2\dot{ H}- 3{H}^{2}-\frac{n-1}{R}(2H\dot {R}+\ddot {R})\nonumber\\
-\frac{(n-1)(n-2){\dot {R}}^{2}}{R^2}-\frac{(n-1)R}{2n}
\end{equation}
Now $\dot {H}=\frac{-u}{t^{2}}$.
Therefore 
$R=6[\dot {H} + 2 {H}^{2}]=\frac{6u(2u-1)}{t^{2}}$ ;
$\dot {R}=-\frac{12u(2u-1)}{t^{3}}$ and $\ddot {R}=\frac{36u(2u-1)}{t^{4}}$.
Using these expressions in (48) we get  
\begin{equation}
p=\frac{-36n^{4}+84n^{3}-61n^{2}+14n-1}{(n^{2}+4n+4)t^{2}} = \frac{v}{t^{2}}
\end{equation}
where $v=\frac{-36n^{4}+84n^{3}-61n^{2}+14n-1}{(n^{2}+4n+4)}$. 
Alternatively,  the $k-$ essence lagrangian (pressure) (22) 
gives 
\begin{equation}
L=-c_{1}\frac{u^{2}}{t^{2}} - c_{2}V\frac{d\phi}{dt}t^{-3u}
\end{equation}
Equating (49) and (50) and solving for $\phi(t)$ gives 
\begin{equation}
\phi(t)= A- \frac{(v+c_{1}u^{2})}{c_{2}V(3u-1)}t^{(3u-1)} 
= A - B t^{\frac{-6n^{2}+8n-1}{2+n}}
\end{equation}
where $A$ is a constant of integration and
$B= \frac{(v+c_{1}u^{2})}{c_{2}V(3u-1)}$. 
Note that we can write with $\beta= \frac{-6n^{2}+8n-1}{2+n}$,  
\begin{equation}
\phi(t)= A - B t^{\beta}
= A - B e^{ln~ t^{\beta}}\nonumber\\
= A - B e^{\beta~ ln~ t}\nonumber\\
\sim A - B e^{\beta (t-1)}
\sim (A - B -B\beta) - B\beta t\nonumber\\
\end{equation}
or
\begin{equation}
\phi(t)= constant - B\beta t
\end{equation}
So $\phi$ at small times is again like an inflationary field \cite{R9}.

\section{Conclusion}

The importance of this work is that Einstein gravity and certain $f(R)$ gravity theories, coupled to the same non-canonical $k-$essence lagrangian, 
can lead to similar cosmological scenarios when a certain scaling relation 
involving the $k-$essence fields is  satisfied in both regimes. So although the underlying gravity theories are different, the cosmological scenarios are realistically 
similar. This observation is in tune with other branches in physics where existence 
of scaling relations signify various genre of universality for theories with different 
microscopic lagrangians. Moreover, as the said scaling relation involves the derivatives of the the dark energy scalar fields only, it may be conjectured that a 
major contribution to the cosmological consequences come from the $k-$essence fields.
     
In this work, we first discuss a modified gravity theory with $f(R)=R^2$ coupled to a non-canonical dark energy lagrangian  $L=-V(\phi)F(X)$, with $X=\nabla_{\mu}\phi\nabla^{\mu}\phi$ and $V(\phi)$ a constant. The modified Friedman equations are subjected to the constraint $X (\frac{dF}{dX})^{2}=Ca(t)^{-6}$ and solutions obtained for the 
scale factor $a(t)$. The deceleration parameter $Q$ and the $k-$essence scalar field $\phi$ are then determined. The following cosmological scenarios are obtained:\\ 
(a) Radiation dominated Ricci flat universe with deceleration parameter $Q=1$.The scalar field takes the form of an inflaton field  for for small times.\\ (b) The deceleration parameter is always negative and we have accelerated expansion of the universe right from the beginning of time. Here also the scalar field is similar to an  inflaton field for small times.\\ (c)The deceleration parameter $Q= -5$ and again we have an accelerated expansion of the universe with the scalar field akin to an inflaton field for small 
times.\\ (d)A generalisation to $f(R)= R^n$is then discussed. It is shown that whenever $n > 1.780$ or $n < - 0.280$ , $Q$ will be negative and we will have accelerated expansion of the universe. At small times the scalar field again behaves like an inflaton.

\end{document}